# Modular Momentum of the Aharonov—Bohm Effect on Noncommutative Lattices

Takeo Miura

Advanced Institute for Mathematical Science,Ltd

1-24-4 Motojyuku,Higashimatuyama,Saitama 355-0063,Japan

Email : t_miura@msg.biglobe.ne.jp

Abstract

Based on the technique of noncommutative geometry, it is shown that, by means of the concept of $\theta$-quantization, there is an equivalence between the notion of the modular momentum of the Aharonov-Bohm effect and the notion of a noncommutative lattice over a circle poset.

Ⅰ. Introduction

The Aharonov-Bohm effect's modular momentum on noncommutative lattices shall be constructed. We have studied the construction of fields over the Penrose Tiling(The Penrose Lattices) for a plane[1]. As a very simple example of a quantum mechanical system ,the $\theta$–quantization of the wave function for a particle on a lattice is shown in [3]. We noticed that by means of $\theta$–quantization, the theoretical structure of the Aharonov-Bohm effect's modular momentum is equivalent to the theory of the above example[1].

In this article, the equivalence of the both above theories shall be described concretely, such as the theoretical structure of the Aharonov-Bohm effect can be studied with the techniques of non-commutative geometry on non-commutative lattices.

For the self-consistency of the explanation in the article, we make reference to the summary of [2] for the modular momentum and also to the summary of [3] for the $\theta$–quantization of wave functions on a lattice for the circle.                .

Ⅱ. The Aharonov-Bohm effect and Quantum nonlocality[2]

Ⅱ-1. Consider electrons of wavelength $\lambda$ diffracting through a heavy grating, the grating consists of narrow slits.

1.The electrons scatter into discrete directions defined by angles $\theta_n$ (n : an integer) :$\sin\theta_n = n\lambda/\ell$, in these directions the partial electron waves interfere constructively. These diffraction satisfy the following conditions where the n slits are spaced a distance $\ell$ apart, and because of the grating is heavy, the energy of the electrons is practically the same before and after diffraction : hence their momentum p = $\hbar/\lambda$ and wavelength remain the same, according to the de Broglie relation.

2. The transverse momentum $p_x$ of an electron scattered through an angle $\theta_n$ is $p_x$ = $p\sin\theta_n$ = $n\hbar/\ell$. These satisfy the following conditions where the incident electrons wave move parallel to the y-axis and diffract in the xy-plane, and the grating is free to move in the x-direction, the x-component of momentum is conserved during diffraction, and the grating acquires momentum $nh/\ell$ in the x-direction from the electron. The electron and grating exchange transverse momentum only in multiples of $h/\ell$.

Ⅱ-2. Next, we place a solenoid between neighboring slits, the solenoids must move independently of the grating.

1. If all the solenoids carry the same flux $\Phi_B$, then electrons, after passing the grating, will scatter into a new set of angles $\theta'_n$ defined by $\sin\theta'_n = (n + (e\Phi_B/2\pi\hbar c))(\lambda/\ell)$. These satisfy the following condition where a solenoid carrying a flux $\Phi_B$ contributes $e\Phi_B/\hbar c$ to the relative phase of partial waves passing on either side of it.

2. The constructive interference now corresponds to a change in the electron's transverse momentum of $p_x = p \sin\theta'_n = (n+(1/2))(h/\ell)$. This satisfies the following condition where if the extra phase due to the solenoids is $(e\Phi_B/\hbar c) = \pi$, then the pattern of lines of constructive interference will be shifted by half the separation between neighboring lines.

Ⅱ-3. We can arrange for the electric and magnetic fields of each solenoid to vanish wherever the electrons go. Thus the electrons and solenoids must exchange momentum non-locally. Nonlocal exchange of momentum is apparently an unobservable effect, for details, see[2].

Ⅱ-4. Modular momentum[2]

The operator $\exp(ip\ell/\hbar)$ reveals the relative phase $\alpha$, a nonlocal aspect of $\varphi_\alpha$ ($\varphi_\alpha = \varphi_1 + e^{i\alpha}\varphi_2$, this equation is interference wave equation), because it translates the wave function. If we replace p by $p - nh/\ell$ in $\exp(ip\ell/\hbar)$, the operator remains invariant, since $\exp(i2\pi n)=1$. So $\exp(ip\ell/\hbar)$ does not depend on all of p; it depends only on p mod $nh/\ell$.

Thus, this quantity is called the modular momentum

$$p \bmod p_0, \quad p = p_1 + np_0, \quad 0 \leq p_1 \leq p_0, \quad p_0 = h/\ell.$$

then, $2\pi np_0 \leq 2\pi p \leq 2\pi(n+1)p_0$. This modular momentum is a circle length $2\pi p_0$.

Ⅲ. Line bundles on a circle poset and $\theta$-quantization[3]

The modular momentum of the Aharonov-Bohm effect can be studied with techniques of noncommutative geometry on noncommutative lattices. We shall construct the $\theta$-quantization of a particle on a lattice for a circle. We shall do so by constructing an appropriate 'line bundle' with a connection.

The real line $R^1$ is the universal covering space of the circle $S^1$, and the fundamental group $\pi_1(S^1) = Z$ acts on $R^1$ by translation.

$$R^1 \ni x \to x + N, \quad Z \ni N,$$

The quotient space of this action is $S^1$ and the projection : $R^1 \to S^1$ is given by $x \to \exp(i2\pi x)$. Now, the domain of a typical Hamiltonian for a particle on $S^1$ need not consist of functions on $S^1$. Rather it can be obtained from functions $\varphi_\theta$ on $R^1$ transforming under an irreducible representation of $\pi_1(S^1) = Z$. According to $\varphi_\theta(x + N) = \exp(iN\theta)\varphi_\theta(x)$, projection is $\rho_\theta : N \to \exp(iN\theta)$.

One obtains a distinct quantization, called $\theta$-quantization, for each choice of $\exp(i\theta)$. Equivalently, wave functions can be single valued functions on $S^1$ while adding a 'gauge potential' term to the Hamiltonian. To be more precise, one constructs a line bundle over $S^1$ with a connection one-form given by $i\theta dx$.

Ⅳ. $\theta$-quantization($p_0$-quantization) of Modular momentum

Now, we give an equivalence between modular momentum and circle variables, this equivalence is a simple key idea of our research. We showed in Ⅱ-4 that the modular momentum of the Aharonov-Bohm effect could be represented over the lattices on a circle. The identification of both representation are attributed to replacements by variables as follows :

$$\theta \to \hbar/\ell = p_0, \quad x = x_1 + N \to p/p_0 = p_1/p_0 + n,$$

$\varphi_{p_0}$ on p was transformed by irreducible representation $\rho_{p_0}$,

$$\rho_{p_0} : p/p_0 \to \varphi_{p_0}(p/p_0), \quad \rho_{p_0} : (p_1/p_0) + n \to \exp(inp_0) \cdot \varphi_{p_0}(p_1/p_0)$$

Ⅴ. A representation of the noncommutative lattices on a circle poset

Ⅴ-1. Quantization of noncommutative lattices on a circle[3]

The algebra $A$ of a noncommutative lattice on a circle, as it is AF(approximate finite dimensional), can be approximated by algebras of matrices. The simplest approximation is a commutative algebra $C(A)$, this is $P_{2N}(S^1)$, of the form $C(A) = C^N = \{c=(\lambda_1, \lambda_2, \cdots, \lambda_N) : \lambda_i \in C\}$.

$P_{2N}(S^1)$ can produce a noncommutative lattice with 2N points by considering particular class of not necessarily irreducible representations. The additional condition is N =

N+1. The partial order, or equivalently the topology, is determined by the inclusion of the corresponding kernels as in [3].

We have, over $C(A)$, a K-Cycle$(\mathcal{H}, D)$, and for $\mathcal{H}$(The Hilbert space), we take $\mathbb{C}^N$, on which we represent elements of $C(A)$ as diagonal matrices.

$C(A) \ni c \mapsto \operatorname{diag}(\lambda_1, \lambda_2, \cdots, \lambda_N) \in \mathcal{B}(\mathbb{C}^N) = M_N(\mathbb{C})$. Elements of sections $\mathcal{E}$ will be realized in the same manner, $\mathcal{E} \ni \eta \mapsto \operatorname{diag}(\lambda_1, \lambda_2, \cdots, \lambda_N) \in \mathcal{B}(\mathbb{C}^N) = M_N(\mathbb{C})$.

We need a Laplacian $\Delta$ and a potential W for a quantum theory, as a Hamiltonian can be constructed from these ingredients. In order to define a Laplacian, we need a self-adjoint operator $D$ to define the 'exterior derivative' d and a matrix of one-form $\rho$ as a connection.

By identifying N+j with j, we take for the operator $D$, the N×N self-adjoint matrix with elements

$$D_{ij} = (1/\varepsilon\sqrt{2})(m^* \delta_{i+1,j} + m \delta_{i,j+1}), \quad i,j = 1,\cdots,N.$$

As for the connection 1-form $\rho$ on the bundle $\mathcal{E}$, we take it to be the hermitian matrix with elements

$$\rho_{ij} = (1/\varepsilon\sqrt{2})(\sigma^* m^* \delta_{i+1,j} + \sigma m \delta_{i,j+1}), \quad i,j = 1,\cdots,N. \quad mm^* = 1,$$

$$\sigma = \exp(-i\theta) - 1, \quad i,j = 1,\cdots,N$$

The curvature of $\rho$ vanishes, $d\rho + \rho^2 = 0$. It is also possible to prove that $\rho$ is a 'pure gauge' for $\theta = 2\pi k$, with k any integer, there exists $\rho = c^{-1} dc$. If $c = \operatorname{diag}(\lambda_1, \lambda_2, \cdots, \lambda_N)$, then any such c will be given by $\lambda_1 = \lambda, \lambda_2 = \exp(i\theta/N) \cdot \lambda, \cdots, \lambda_j = \exp(i\theta(j-1)/N) \cdot \lambda, \cdots, \lambda_N = \exp(i\theta(N-1)/N) \cdot \lambda$, with $\lambda \neq 0$. $\lambda$ is the eigenvalue of Laplacian, $\Delta_\theta \eta = \lambda \eta$, see[3].

V-2. The generation of the Weyl form of commutation relation[1],[4]

We get the Weyl form of matrix operators with AF-algebra on the above $C(A)$. c is the diagonal N×N matrix with the above eigenvalues, and $D$ can be separated as $D = D_1 + D_1^*$.

c is a multiplicative operator and $D_1$ is a shift operator. We get the Weyl form of commutation relation,

$$c \cdot D_1 = \exp(-i\theta/N) \cdot D_1 \cdot c, \qquad \theta = \hbar/\ell,$$

By the existence of this equation, the construction of the covering, (as noncommutative poset), over $S^1$-space can be proved.

VI. Conclusion

We showed that the modular momentum of the Aharonov-Bohm effect could be constructed on non-commutative lattices.
The non-commutative lattices imply the the discreteness of space. Recently, at Fermi Laboratory(in USA), an experiment for the assurance of the discreteness of 'space and time' has been preparing[5].